\documentclass[preprint]{revtex4}
\usepackage{graphicx}
\usepackage{epsfig}
\input{epsf}

\begin{document}

\title{Evidence of Conformational Changes in Adsorbed Lysozyme Molecule on Silver Colloids}
\author{Goutam Chandra$^1$, Kalyan S. Ghosh$^2$, Swagata Dasgupta$^2$ and Anushree Roy$^1$}
 \affiliation{$^1$Department of Physics,
Indian Institute of
Technology, Kharagpur 721302, India\\
$^2$Department of Chemistry, Indian Institute of Technology,
Kharagpur 721302, India}

\begin{abstract}
In this article, we discuss metal-protein interactions in the
Ag-lysozyme complex by spectroscopic measurements. 
The analysis of the variation in relative intensities of SERS bands
reveal the orientation and the change in conformation of the protein
molecules on the Ag surface with time. The interaction kinetics of
metal-protein complexes has been analyzed over a period of three
hours via both Raman and absorption measurements. Our analysis
indicates that the Ag nanoparticles most likely interact with
Trp-123 which is in close
proximity to Phe-34 of the lysozyme molecule.\\


\end{abstract}

\maketitle

\section{Introduction}
Metal nanoparticles are finding increasing use in nanomedicine for
surface interactions with an array of proteins and/or small
molecules. Metal nanoparticles tagged to biomacromolecules are
oftentimes used to identify specific antibody-antigen binding sites
in cells and tissues \cite{Horisberger,Roth}. Surface Enhanced Raman
scattering (SERS) is an extremely sensitive technique for monitoring
the adsorption of species of very low concentration and for
characterizing the structure and orientation of the adsorbed species
on the rough metal surface or metal colloidal particles
\cite{Kneipp}. Nonresonance SERS is primarily sensitive to the
species attached to the metal surface or present within a few Å of
the metal-dielectric interface as well as to the orientation of the
ad-molecule \cite{Haynes}. Thus, it is possible to characterize
molecular rearrangement on the metal surface by SERS mechanism. This
implies that conformational changes occurring in the protein
structure or due to denaturation through possible interactions with
metal particles can be detected from a detailed analysis of the SERS
spectrum. Modification of the specific interaction sites by covalent
or non-covalent methods makes the possibilities of study enormous.
In this regard, the interactions of silver nanoparticles with
several proteins have been investigated. Some of the proteins
studied with silver (Ag) colloids are hemoglobin  \cite{Xu},
antirabbit IgG \cite{Grabbe}, albumin \cite{Ahren} and lysozyme
\cite{Hu,Podstawka,Chumanov}.

Lysozyme is a small monomeric globular protein consisting of 129
amino acids that has the ability to disrupt many bacterial functions
including their membrane structure
\cite{Pellegrini,Ibrahim,Pellegrini2,Croguennec}. The protein has
$\alpha$-helix and $\beta$-sheet components with four disulfide
bonds. It contains six tryptophan (Trp) residues, three of them are
located at the substrate binding sites, two in the hydrophobic
matrix box, while one is separated from the others \cite{Sheng}.
Among them Trp 62 and Trp108 are the most dominant fluorophores,
both being located at the substrate binding sites \cite{Imoto}.
Lysozyme action on bacteria has been the focus of a lot of research,
but interaction studies with Ag colloids have not previously
revealed the specific interaction sites on the protein. We have
further probed the interactions of lysozyme with Ag nanoparticles to
obtain an insight into the specific residues involved in the
interaction. This has been investigated by spectral methods and new
insights have been obtained in the understanding of how this protein
is involved in the interaction with Ag colloids.

Earlier studies on the SERS spectrum of lysozyme molecules adsorbed
on Ag electrodes and hydrosols indicate the proximity of the S-S
bond and aromatic amino acid residues of lysozyme molecule on the Ag
surface \cite{Hu,Podstawka,Chumanov}. These articles report the
orientation of the aromatic amino acid side-chains located on the
surface of the metal electrodes. The role of an $\alpha$-helical
component of the lysozyme molecule for binding on the metal surface
has been shown  \cite{Hu,Podstawka} and a recent study has reported
the conformational changes in lysozyme molecule with temperature,
while adsorbed on a gold nano-patterned SERS substrate \cite{Das}.
Thus previous studies indicate the possible regions of interaction
of the lysozyme molecule while interacting on the surface of metal
colloids. It is necessary to probe the specific region of binding of
the protein molecule to the Ag colloidal surface to permit an
understanding of which regions of the protein would be free to
interact with ligands and/or drugs. For example, the antimicrobial
macromolecule formed by a lysozyme-polyphenol complex can be used as
a carrier for phenolic antioxidants. Epigallocatechin gallate (EGCG)
\cite{Ghosh} and Triclosan \cite{Hoq} molecules are known to form
complexes with lysozyme via Trp-62 and Trp-63 residues. The
knowledge of the specific binding site in Ag-lysozyme complex can
provide guidelines to develop nanoparticle tagged lysozyme molecule
for drug delivery. It may be noted that the lysozyme is a potential
carrier molecule as a renal-selective drug carrier for delivery of
the angiotensin-converting enzyme (ACE) inhibitor captopril
\cite{Jaiprakash}.

\section{Materials and Methods}

\noindent \textbf{Synthesis route  for preparing stabilized
particles:} Ag nanoparticles have been synthesized by chemical
routes, using sodium borohydride (NaBH$_{4}$) as reducing agents.
Silver nitrate (AgNO$_{3}$), sodium borohydride (NaBH$_{4}$) of
analytical reagent grade (SRL, India), were used to prepare the Ag
sol.  A colloidal silver solution was prepared in deionized water
following the method described by Creighton et al.
\cite{Creighton:1979}. Essentially, in this chemical route,
AgNO$_{3}$ is reduced by an excess amount of NaBH$_{4}$. 2.2$\times$
10$^{-3}$ M AgNO$_{3}$ was added dropwise to 1 mM NaBH$_{4}$ at 4
$^\circ$C. Stirring for 20 minutes was necessary to stabilize the
colloidal solution. Later, it was left at room temperature for
approximately 1 hour. The excess NaBH$_{4}$ evaporated and the
remaining solution became transparent yellow in color. Colloids
generated by above method were stable over a considerable period of
time (4-6 weeks). However, all our reported results are either with
freshly prepared Ag sol or aged only for a day.

Chicken egg-white lysozyme was obtained from Sigma-Aldrich. In
solution, depending on the pH values, the majority of lysozyme
molecules assume different structural forms. The protein has a zero
net charge at the isoelectric point (pI). Interactions of neutral
protein molecules with water is less favorable than when they are in
the charged state. At the pI, the solubility of proteins in water is
low and the repulsion between protein molecules is also a minimum.
These two factors help in adsorption of  the molecules on metal
surfaces for pH of the solution close to pI. For lysozyme the value
of pI is 10.5. At pH 10.5 the protein is not soluble in water, which
would cause problem in comparing the SERS spectrum and normal Raman
spectrum of lysozyme molecule. At pH 11 the molecule undergoes rapid
hydrolysis \cite{Lord}. Moreover, the pH of untreated metal sol was
between 7.0-7.5. For these reasons we have maintained the pH of the
lysozyme solution in slightly acidic state, such that the effective
pH in the final sol (lysozyme+Ag sol) is close to 7.0-7.5 in all our
experiments.  4 mM solution of lysozyme was prepared in deionized
water. For SERS measurement the volume ratio of Ag sol to protein
solution was maintained at 1:1.

SERS spectra were measured using a 488 nm Argon ion laser as an
excitation source using a microRaman spectrometer with 100X
objective lens. The spectrometer is equipped with 1200 g$/$mm
holographic grating, a holographic super-notch filter, and a Peltier
cooled CCD detector. The laser power on the samples was 5 mW. The
unchanged circular dichroism spectra of laser irradiated and
un-irradiated samples eliminate the possibility of
damage/conformational changes in the protein structure by laser
irradiation under the given experimental conditions. The data
acquisition time for each Raman and SERS spectrum was 120 sec.


UV-Visible spectra were measured by Spectrascan UV 2600 (Chemito).
To study interaction kinetics, measurements were carried out
immediately after addition of lysozyme in Ag sol and subsequently
after 15 mins, 30 mins, 1 hr, 2hrs and 3 hrs.

For optical absorption measurements the samples were kept in a
microcuvette with optical pathlength 1 cm. For SERS measurements a
drop of solution was taken on a glass slide and allowed to dry. It
took around 10-15 minutes to dry the samples. During kinetic
measurements, to investigate the interaction between metal colloids
and protein molecules in the solution, drops were deposited after
the specific time intervals as given above. Thus, we carried out all
measurements after the completion of metal-molecule interaction for
the required time, in ambient conditions.
All measurements (SERS and optical absorption) and analysis have
been carried out  in triplicate.

PyMol \cite{Pymol} was used for visualization of the protein
conformation. The accessible surface area (ASA) of all the amino
acid residues in uncomplexed lysozyme are estimated using NACCESS
\cite{ASA}.

\section{Effect of added lysozyme molecules on metal colloids}
\subsection{Analysis of Plasmon Band of Silver Colloids} All biomolecules do not interact with
metal colloids in the same way. Some of them result in agglomeration
of the colloidal particles, others do not cause aggregation but
modify the surface charge of the particle \cite{Ahren}. To
understand the effect of added lysozyme molecules on Ag colloids, we
study the plasmon band of Ag particles in the sol on adsorption of
the protein molecules. Plasmon band of Ag colloids is shown by black
dotted line in Fig. 1 (a). Ag sols absorb light at 398 nm, due to
the dipolar surface plasmon of small spherical particles of Ag. The
average particle size of Ag has been estimated to be $\sim$ 100 \AA.


Time-dependance of the plasmon band of Ag colloids in the sols is
shown in Fig. 1 (a). Spectra, recorded just after the addition of
protein molecules in the metal sols, are shown by black solid line.
Subsequent changes in the spectra with time (after 15 mins., 30
mins, 1 hr, 2 hrs and 3 hrs.) are shown by colored solid lines in
the Fig. 1 (a). Addition of lysozyme to the sol results in (i)
decrease in intensity and (ii) broadening with a (iii) red shift of
the plasmon resonance band of Ag particles in the sol. The decrease
in relative intensity of the plasmon absorption band in solutions
(ratio of maximum intensity of the plasmon band of Ag colloids after
addition of protein to that of the  Ag colloids), is shown in Fig.
1(b). We observe that the intensity of the plasmon band decays
exponentially. The decay time constant of intensity is estimated to
be 60 min in the sol.  The observed broadening and the red shift of
the plasmon band indicate formation of bigger particles in the sol
by added lysozyme molecules. Furthermore, the decrease in the
intensity of the peak can be explained by precipitation of metal
particles from the sol followed by immediate aggregation.





\section{Effect of metal colloids on added lysozyme molecules}
\noindent \textbf{Analysis of SERS Spectra}


Next, we focus on the effect of metal particles on added lysozyme
molecules in the sol.  Raman spectra and SERS spectra of 4 mM
lysozyme in water and in the Ag sol (just after addition) are shown
by solid lines and dotted lines respectively in Fig. \ref{SERSP}.
Appearance of the characteristic feature of the Ag-N vibrational
mode at 233 cm$^{-1}$ \cite{Hu} in the SERS spectrum (shown in the
inset of Fig. \ref{SERSP}) is a clear signature of metal-protein
interaction in the sol via N atom of the ad-molecule. Time variation
of SERS spectra of lysozyme in Ag sol over the spectral range
400--900 cm$^{-1}$, 900--1100 cm$^{-1}$ and 1100--1700 cm$^{-1}$ are
shown in Fig. \ref{SERSL}(a)--(c)
The assignments of the prominent bands (indicated by $\star$ marks
in Fig. \ref{SERSP}) are summarized in Table \ref{SERST}.
To study the interaction of the Ag-lysozyme complex, we have
estimated the relative intensity of each feature in the range over
400--1100 cm$^{-1}$ [Fig. \ref{SERSL}(a) and (b)] with respect to
the intensity of the peak at 877 cm$^{-1}$(vibration of N$_1$H site
of Trp amino acid residues in lysozyme). The methylene
$\delta$(CH$_2$) vibration at 1442 cm$^{-1}$ has been used as
reference for the spectral features in the range 1100--1700
cm$^{-1}$ [Fig. \ref{SERSL}(c)]. Both these peaks (at 877 and 1442
cm$^{-1}$) maintained a constant frequency and were of quite strong
intensities. Invariant Raman shift of a band indicates unchanged
chemical bonding responsible for these particular vibrational modes
over experimental duration.



It was interesting to note that the relative intensity of all
vibrational bands of the adsorbed lysozyme molecule do not follow
the same trend.  A possible explanation for the time-evolution of
the spectra can be the conformational changes of the ad-molecules
with time, which we discuss  in detail below.

\noindent \textbf{A. Vibrational modes of amide bands}

Vibrations of peptide backbone in proteins, associated with amide I,
mainly involve the  C=O stretch with a small contribution of C-N
stretching and N-H bending. The same for the amide III region arises
predominantly by an in-plane N-H deformation coupled with a
C${_\alpha}$-N stretch.  In lysozyme, amide bands consist of mainly
three structural components --- three stretches of $\alpha$-helix,
antiparallel pleated $\beta$ sheet, in which the polypeptide is
hydrogen-bonded between one region to another via a hairpin turn of
the chain and random coils, which are folded in an irregular way.
Lysozyme is a globular protein with distinctly demarcated
$\alpha$-helix and $\beta$-sheet regions. These secondary structural
units are connected by random coils. The positions and intensities
of the vibrational modes of amides can be used for an empirical
estimation of these secondary structures in proteins.


SERS spectrum of amide I band appears between 1610--1700 cm$^{-1}$ .
The contributions of the three secondary
structures---$\alpha$-helix, $\beta$-sheet and random coil--- are
expected to occur over the spectral windows 1665-1680 cm$^{-1}$,
1610--1632 cm$^{-1}$ and 1636--1644 cm$^{-1}$, respectively
\cite{Dong}.  The amide III band appears over the spectral range
between 1230 and 1310 cm$^{-1}$--- the signatures of $\alpha$-helix,
$\beta$-sheet and random coils are characterized by spectral peaks
over the range 1280--1320 cm$^{-1}$, 1235--1242 cm$^{-1}$ and
1250--1260 cm$^{-1}$, respectively \cite{Lipert}. The region
1254--1260 cm$^{-1}$ has been assigned to $\beta$-turn \cite{Lord2}.


In the measured SERS spectra of lysozyme in Ag sol, the vibrational
modes of amide III  appear at ~1260 cm$^{-1}$,  as a shoulder of the
strong peak at 1331 cm$^{-1}$ of Trp (Fig. \ref{SERSL}). Thus, it is
difficult to extract the subtle spectral features of amide III band
from spectral analysis. To study the correlation between three
secondary structural components of the adsorbed protein molecule,
the broad envelope of the amide I band of each spectrum needs to be
deconvoluted into the contributions from each of them. Such analysis
may also provide information on conformational changes in lysozyme
with time upon adsorption on colloidal surface. The time-variation
of the vibrational spectra of the amide I band are shown in Fig.
\ref{amide1} by + signs. The deconvoluted and net fitted spectra are
shown in Fig. \ref{amide1} by dotted and solid lines. From the
analysis of the spectral profile of amide I band we obtain the
following: (i) the spectral feature over the spectral range between
1610 and 1632 cm$^{-1}$ for $\beta$-sheet is missing in all spectra
and (ii) the relative vibrational intensities of random coils
(I$_R$) to $\alpha$-helix (I$_\alpha$) increases with time (in the
inset of Fig. \ref{amide1}).


\noindent \textbf{B. Vibrational modes of amino acid residues}

As mentioned before, the sharp SERS spectral line at 877 cm$^{-1}$
corresponds to the vibration of N$_1$H site of Trp residues
\cite{Miura}. The frequency of this band depends on the strength of
the H-bond between Trp and other side chain molecule. The same
feature is expected to shift by 8 cm$^{-1}$ for non-H-bonding Trp
derivatives and lowered by 7 cm$^{-1}$ for H-bonded Trp derivatives.
The single sharp peak at 877 cm$^{-1}$ in the Raman and SERS spectra
indicates that in the metal-protein complex the Trp-residues do not
form any new chemical bond directly with the Ag surface or with
other residues. 
The variation in relative intensities of other vibrational modes of
Trp are shown in Fig. \ref{Trp}(a). The intensity of ring breathing
mode of Phe-residue at 1004 cm$^{-1}$ first increases and then
decreases with time (Fig. \ref{Trp}(b)). The change in
$\delta$(CH$_2$) mode of glutamic acid at 1449 cm$^{-1}$ (just next
to $\delta$(CH$_2$) methylene vibrational mode at 1442 cm$^{-1}$) is
shown in Fig. \ref{Trp}(c). In each plot, the solid line is the
guide to the eye.


We did not find any clear signature of Asp and His (histidine), two
other amino residues of lysozyme, in the SERS spectra of the
molecule.

\noindent \textbf{C. Disulfide bond}

There are four pairs of S-S bond in lysozyme. Among these four
pairs,  two pairs of S-S bonds are near the $\alpha$-helix region
and two others are in the $\beta$-sheet region. The strong SERS
spectrum for vibration of S-S bond appears at 505 cm$^{-1}$ just
after addition of lysozyme in the Ag sol. However, the intensity of
this band decreases with time (Fig. \ref{Trp}(d)). To check the
effect of the reducing environment on the disulfide bond, an
additional experiment with lysozyme and NaBH$_4$ was conducted. The
trend in the change of relative intensity of S-S bond (Fig.
\ref{Trp}(e)) indicates that reduction of the disulphide linkage by
BH$_4^-$ (note that the Ag colloidal particles are stabilized by
BH$_4^-$ in NaBH$_4$ stabilized sol) is possible.




\noindent \textbf{Discussion}


Here we report the specific amino acid residues of lysozyme molecule
which are adsorbed on Ag colloids. Appearance of strong vibrational
modes of Trp and Phe residues in the SERS spectra of lysozyme
molecule and the similar trend in the change in relative intensities
of these modes (Fig. \ref{Trp}(a) and Fig. \ref{Trp}(b)) signify
that both these residues are close to the Ag surface and occupying
adjacent sites in the adsorbed molecule. A lysozyme molecule
contains six Trp residues (Fig. \ref{cartoon1}). The accessible
surface area (ASA) available for these residues are shown in Table
II. Three (Trp-62, Trp-63 and Trp-123) of the six Trp residues of a
free lysozyme molecule have ASA more than 50 {\AA}$^2$ and, hence,
are likely to be free to bind with a substrate.  Trp-62 and Trp-63
are exposed at the edge of the active site cleft and known to
interact specifically with a substrate \cite{Lipert}. Trp-123 is
close to the c-end of the amino acid chain of the molecule. Trp-28,
-108, -111 have relatively less ASA and are expected to be burried.
Of all three Phe residues (Phe-3, Phe-34 and Phe-38) of lysozyme
molecule the ring of Phe-34 is very close to Trp-123 and its ASA is
relatively large compared to that of the other two [Fig.
\ref{cartoon1} and Table II]. Thus, it is most likely that the
lysozyme molecules are adsorbed on Ag surface via the red marked
region which includes both Trp-123 and Phe-34 residues of the
molecule [Fig. \ref{cartoon1}].



Next we discuss the observed SERS spectra of amide I band. It is to
be noted that the ratio of the secondary structure contents i.e
$\alpha$ helix:$\beta$ sheet:random coil in free lysozyme is
expected to be 0.54:0.15:1 \cite{Lippert}. Dominant spectral
features of $\alpha$-helical component in Fig. \ref{amide1} indicate
that lysozyme is adsorbed on the colloidal Ag surface either in an
$\alpha$-helical region or in a loop with an $\alpha$-helical
element
\cite{Swagata}. 
The increase in  the intensity ratio (I$_R$ : I$_\alpha$) in the
inset of Fig. \ref{amide1} indicates unfolding of the $\alpha$-helix
or the $\alpha$-helical turn element to a random coil of adsorbed
protein molecule with time. It is interesting to note that the
$\phi$,$\psi$ angles of  Trp-123 in lysozyme are -124.9,-2.5 [red
arrow in Fig. \ref{cartoon1}], that is close to the $\alpha$-helical
region of the Ramachandran plot.

Furthermore, the disulfide bond is expected to be affected by the
change in protein backbone conformation \cite{Munro}. Of all four
S-S bonds, two of them, (S6-S127) and (S30-S115) are very close to
Trp-123. As the turn element, at the arrow in Fig. \ref{cartoon1},
gets affected due to the interaction with the Ag surface,
S6-S127/S30-S115 bond/bonds may break down and this explains the
decrease in the intensity of S-S vibrational mode with time (Fig.
\ref{Trp}(d)). Furthermore, Glu-7 of lysozyme lies very close to the
S6-S127 bond with an ASA corresponding to 82.4 \AA$^{2}$. If the S-S
bond breaks down (Fig. \ref{Trp}(d)), the vibrational mode of Glu-7
is expected to be affected in a similar fashion (Fig. \ref{Trp}(c)).
Moreover, it is to be noted that this particular region of the
molecule corresponds to the C-terminal end of the protein. Hence,
this region of the molecule is more flexible and susceptible to
changes in the environment.


Addition of lysozyme in Ag sol results in an immediate agglomeration
of the Ag particles. 
The initial increase in the relative intensity of vibrational bands
of amino-acid residues of the molecule with time in Fig.
\ref{Trp}(a) and Fig. \ref{Trp}(b) suggest re-orientation of the
residues to obtain a low energy configuration from an initial high
energy state due to initial favorable electrostatic interactions
during change in conformation.  The time constant for the unfolding
of the $\alpha$-helix is also estimated to be 45 min (inset of Fig.
\ref{amide1}).  If we assume only the local electrostatic
interaction between the adsorbates and metal colloids, all our
results suggest that lysozyme molecules interact with Ag colloids
instantaneously; however, it takes longer time for the protein
molecule to be stabilized on the metal surface. In general,
irreversible laws of thermodynamics are involved in adsorption of
proteins on metal surfaces \cite{Eckert}. Various time scales are
involved in metal-molecule interactions. The fast steps are
sometimes reversible (can be probed only by time-resolved
experimental measurements) and the relatively slow steps (of
time-scale varies from seconds to minutes) result in rearrangements
of  protein structure by the surface environment and in many cases,
can be irreversible \cite{Eckert}. We believe that the initial
metal-protein interaction via agglomeration of the particles
involves the fast process mentioned above and the conformational
changes of the lysozyme molecule with time corresponds to the latter
process. All above conjectures, obtained from spectroscopic
measurements, can be further confirmed by studying interaction of
Ag-lysozyme complexes with other ligands that are currently
underway, for which the interaction sites of lysozyme are already
established.

As mentioned earlier that there are few reports in which SERS
spectra of lysozyme molecules on Ag colloidal surface have been
discussed. Prior knowledge available on the interaction of the
Ag-nanoparticles with lysozyme indicates the $\alpha$-helical and
random coil components \cite{Podstawka}. Our experimental approach
identifies the particular Trp residue involved in the interaction.
Besides this, the effect of the interaction on the intensities of
SERS bands of Phe and disulfide bonds confirm the specific site of
the protein involved in the complexation. Conformational changes in
the protein molecule on interaction with Ag nanoparticles indicate
partial unfolding of the $\alpha$-helix constituted of residues
25-35 of which Phe-34 is a member. Interestingly, it is to be noted
that the parts of the protein molecule with $\beta$-sheet comprising
of Trp-62 and Trp-63, known to have higher bio-activity, remain
unaffected by metal-protein complexation.

\noindent{\textbf{Acknowledgements}}\\AR and SDG thank DRDO, India
for financial assistance\\
\newpage
\begin{center}
\end{center}

\newpage

\noindent{\bf Figure Captions}

\noindent\textbf{Figure 1.} (a)Kinetic study of the plasmon band of Ag
colloids and (b)Time variation in relative intensity
of the plasmon band of the Ag upon addition of lysozyme in the sol.


\noindent\textbf{Figure 2.} Comparison between normal Raman spectrum
of 4$\times 10^{-3}$M lysozyme in water (solid lines) and  SERS
spectrum of the same in the Ag sol (dotted lines). Inset of the
figure shows the characteristic feature of Ag-N vibrational mode.

\noindent\textbf{Figure 3.} SERS spectra of Ag-lysozyme complex
taken at different time over the range (a)450-900 cm$^{-1}$, (b)
900-1100 cm$^{-1}$ and (c) 1100-1700 cm$^{-1}$.

\noindent\textbf{Figure 4.} SERS spectrum (+ signs) and deconvoluted
spectra (dashed lines) of the amide I band. Inset of the figure
shows the variation in intensity ratio of random coil to
$\alpha$-helix components of the protein with time.



\noindent\textbf{Figure 5.} Variation in intensity of vibrational
bands of (a)Trp residues (b)Phe residues (c)glutamic acid (d)S-S bonds of
adsorbed lysozyme in Ag sol and (e)S-S bonds of lysozyme in BH$_4$ environment (without metal colloid) with time respectively.


\noindent\textbf{Figure 6.} Cartoon of lysozyme molecule and its
site of interaction with Ag surface (the region within the red mark.
The $\alpha$-helix element which unfolds upon interaction with metal
surface is shown by arrow.

\newpage

\noindent{\bf Table Captions}

\noindent\textbf{Table I.} Assignment of vibrational bands for
lysozyme SERS spectra.

\noindent\textbf{Table II.} Calculated accessible surface area (ASA)
for Trp, Phe and Glu residues in a lysozyme molecule.

\newpage

\begin{figure}
\epsfig{file=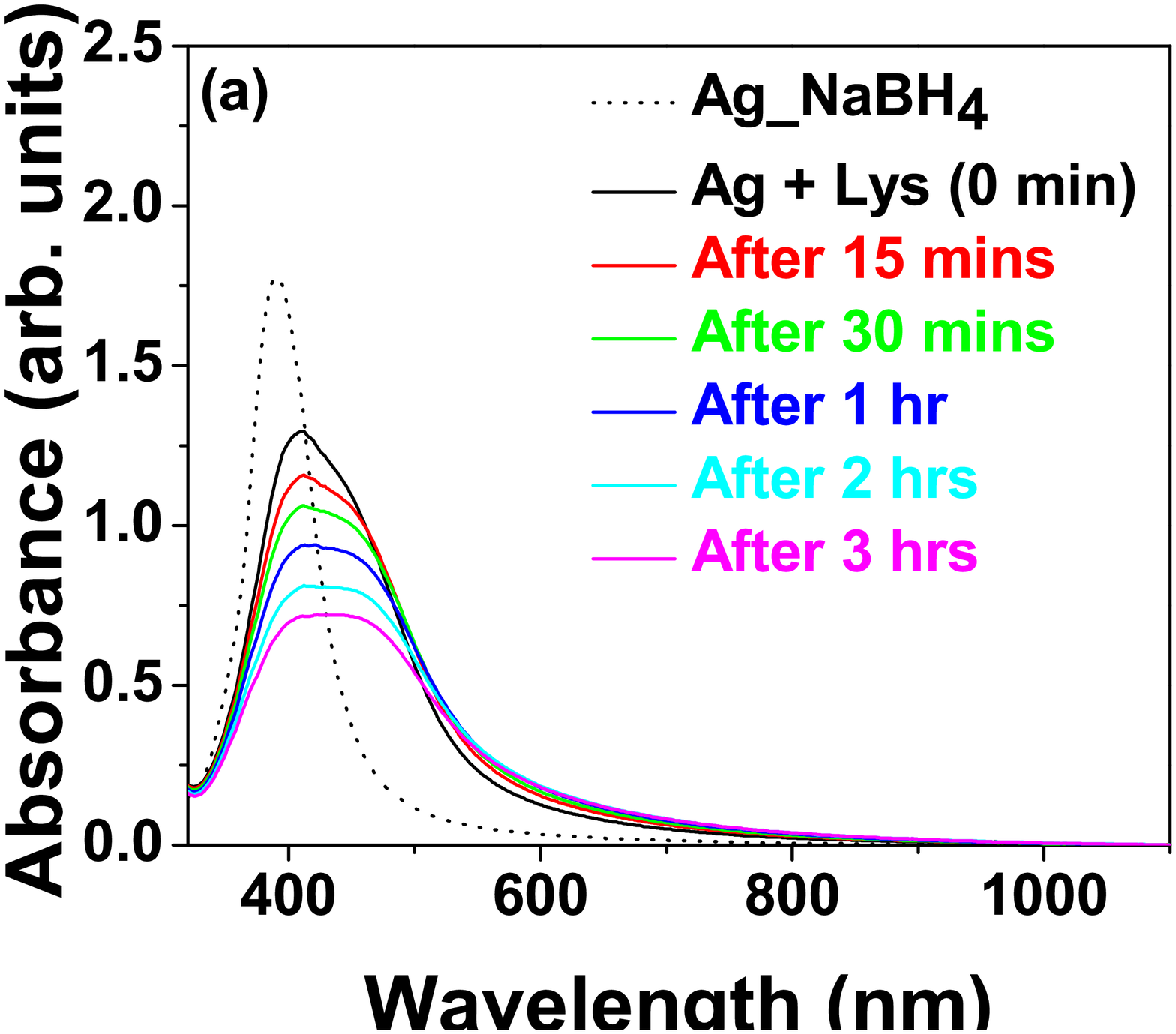,width=2.5in, angle=0}
\epsfig{file=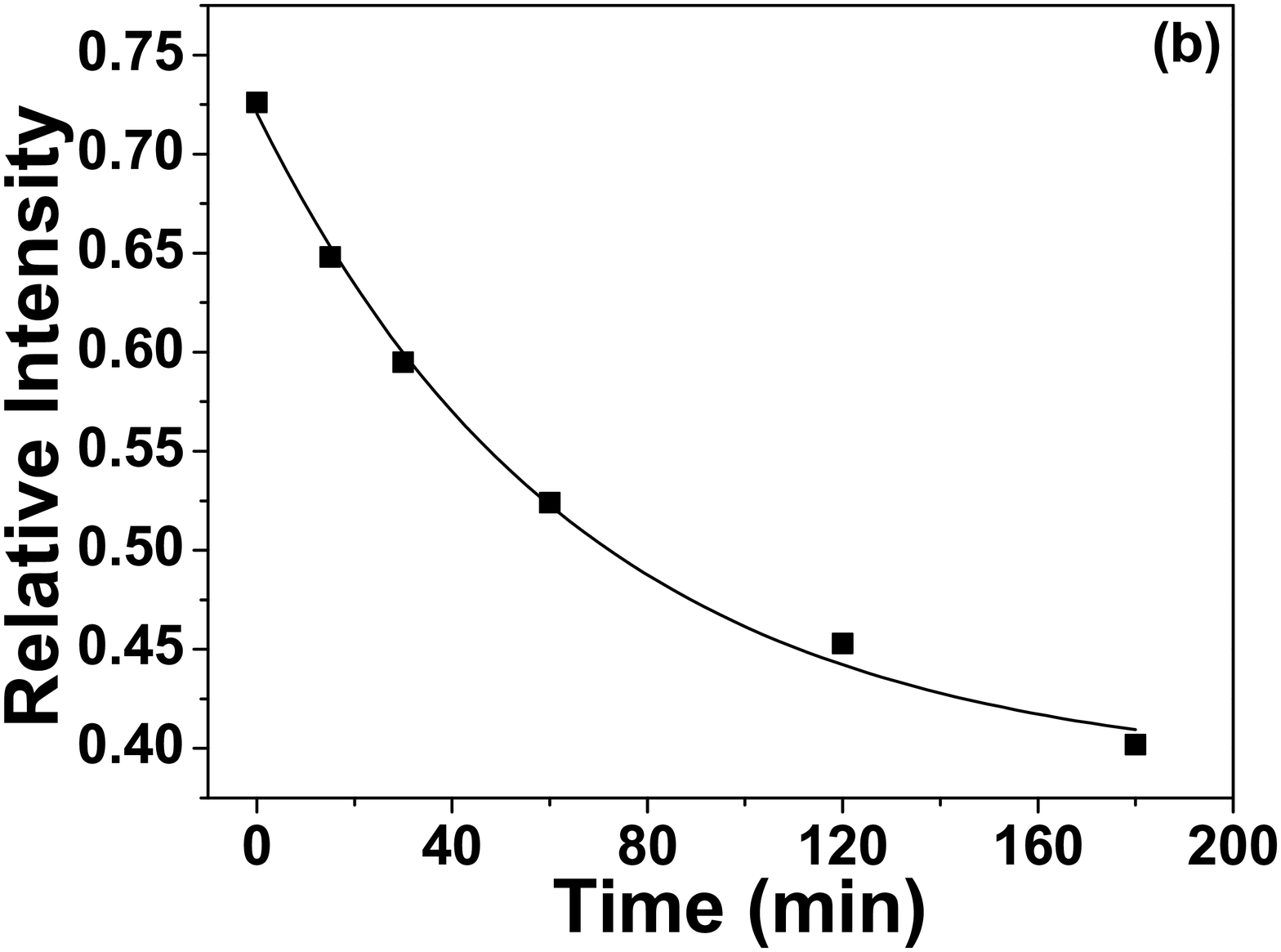,width=2.50in, angle=0}
 \vspace*{0.75in}
\caption{Chandra et al} \label{plastime}
\end{figure}



\newpage

\begin{figure}
\epsfxsize=3.5in\epsffile{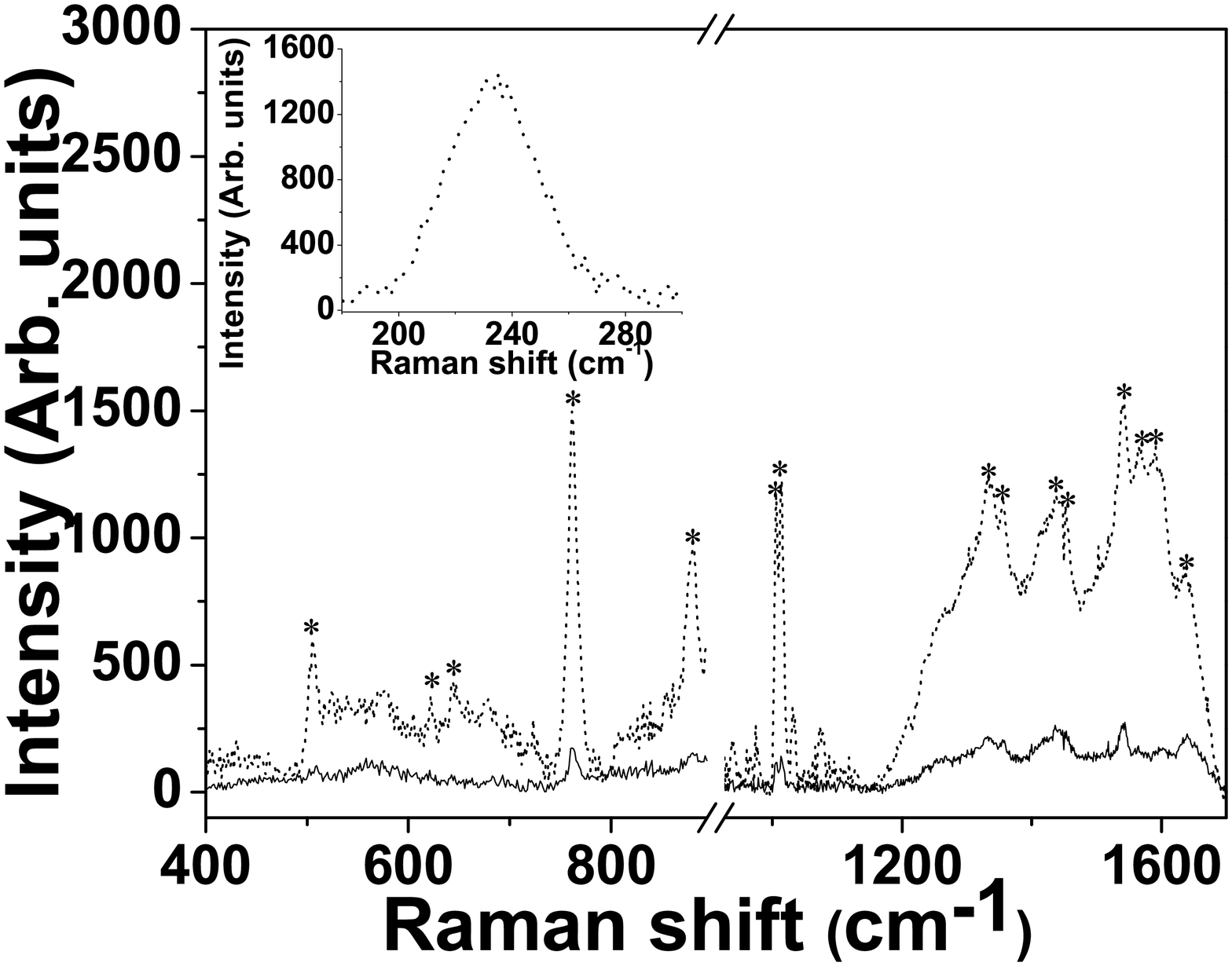}
\caption{Chandra et
al} \label{SERSP}
\end{figure}

\newpage

\begin{figure}
\epsfig{file=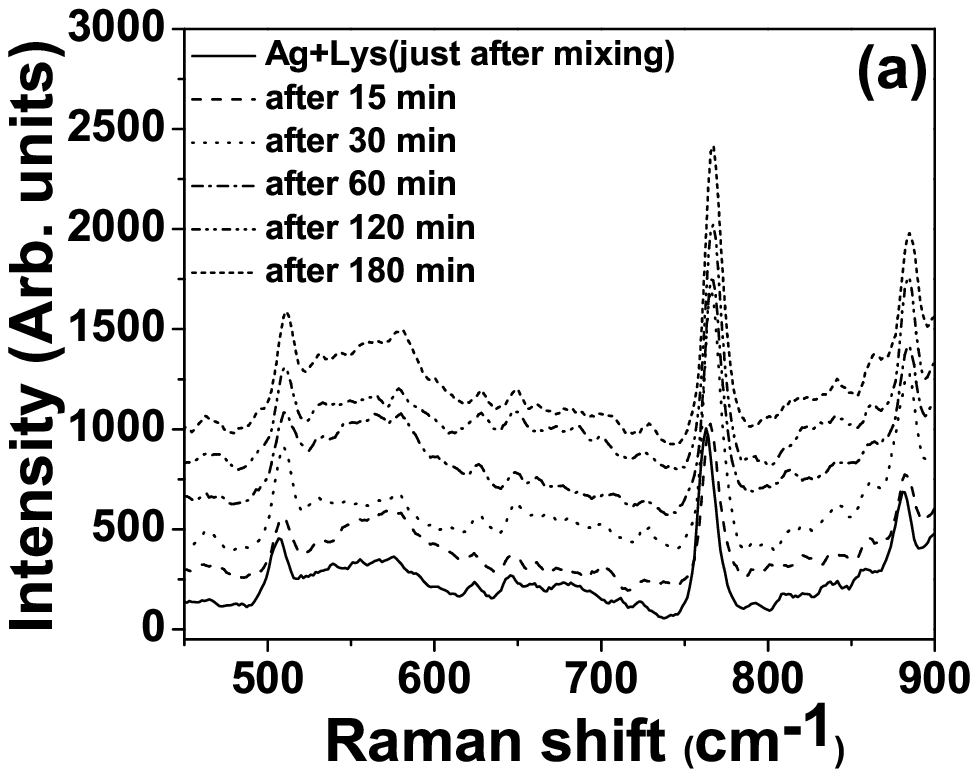,width=2.5in, angle=0}
\epsfig{file=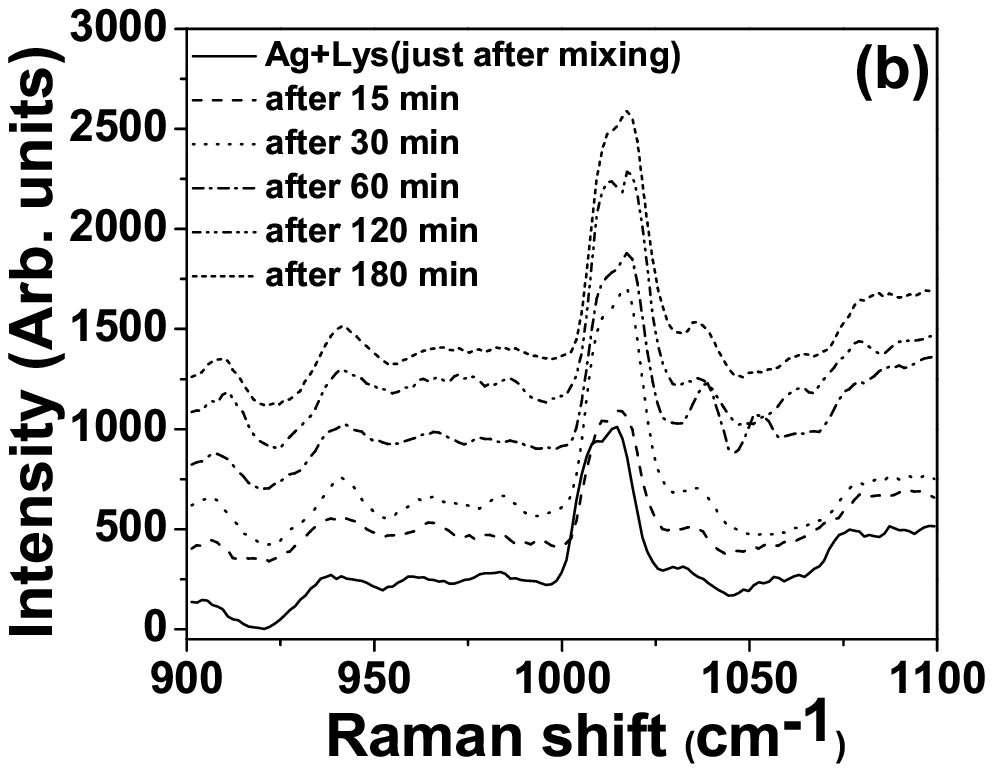,width=2.5in, angle=0}
\epsfig{file=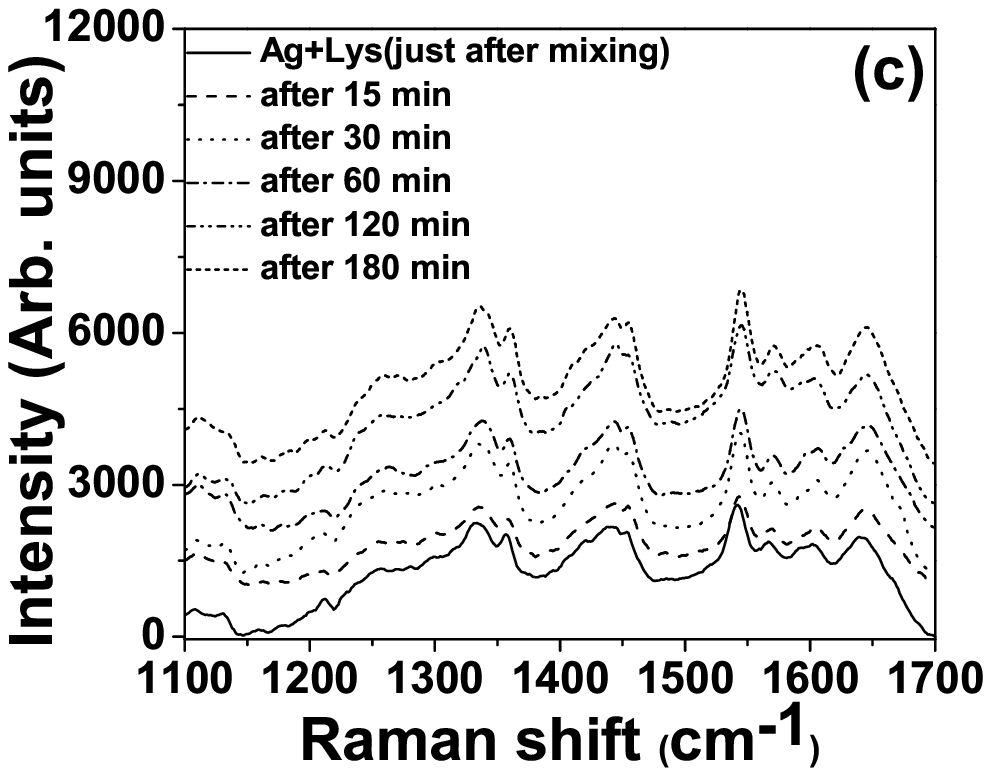,width=2.5in, angle=0} \caption{Chandra et al}
\label{SERSL}
\end{figure}

\newpage

\begin{figure}
\centerline{\epsfxsize=4.0in\epsffile{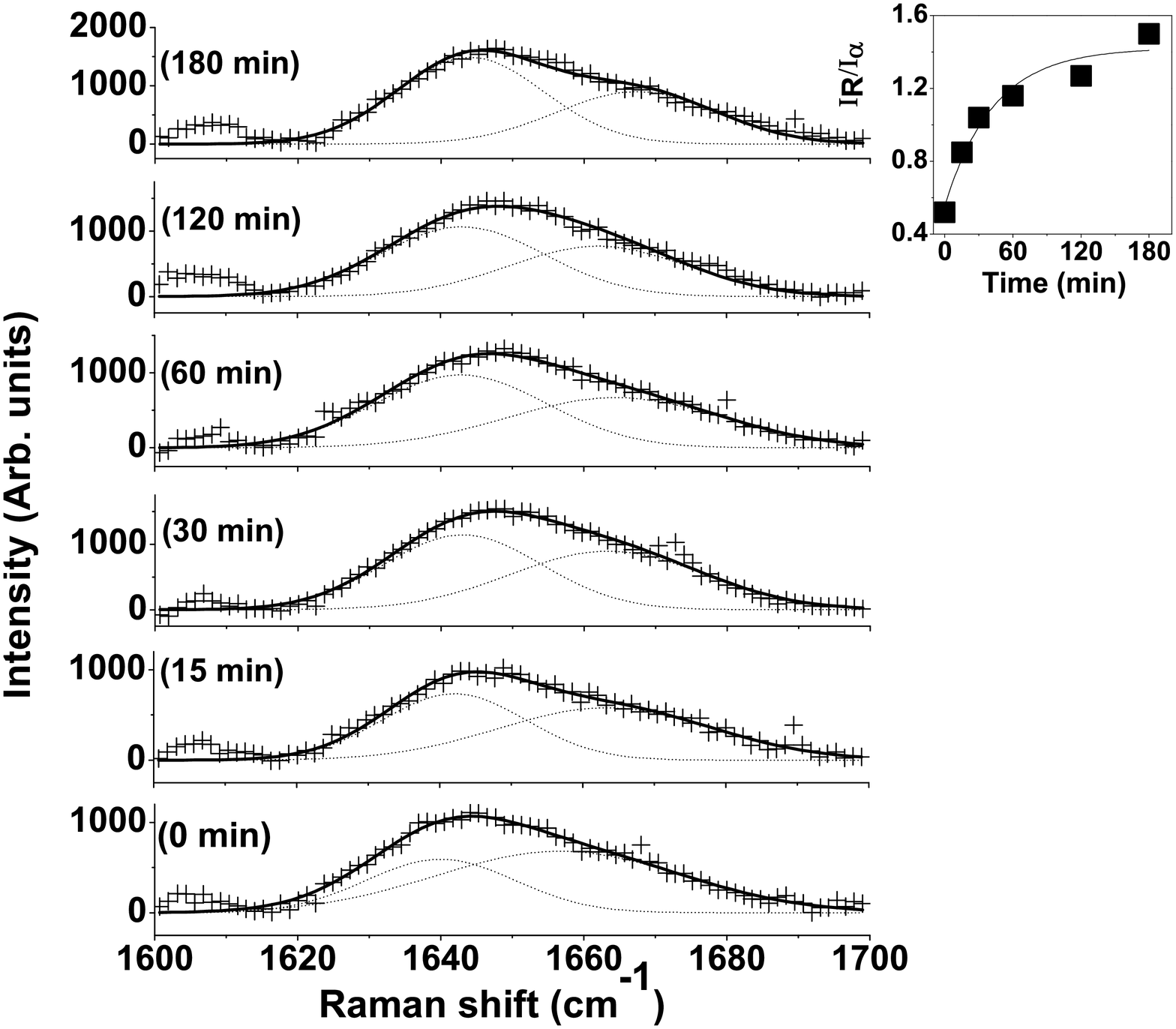}}
\vspace*{0.75in}\caption{Chandra et al} \label{amide1}
\end{figure}



\newpage

\begin{figure}
\centerline{\epsfxsize=3.6in\epsffile{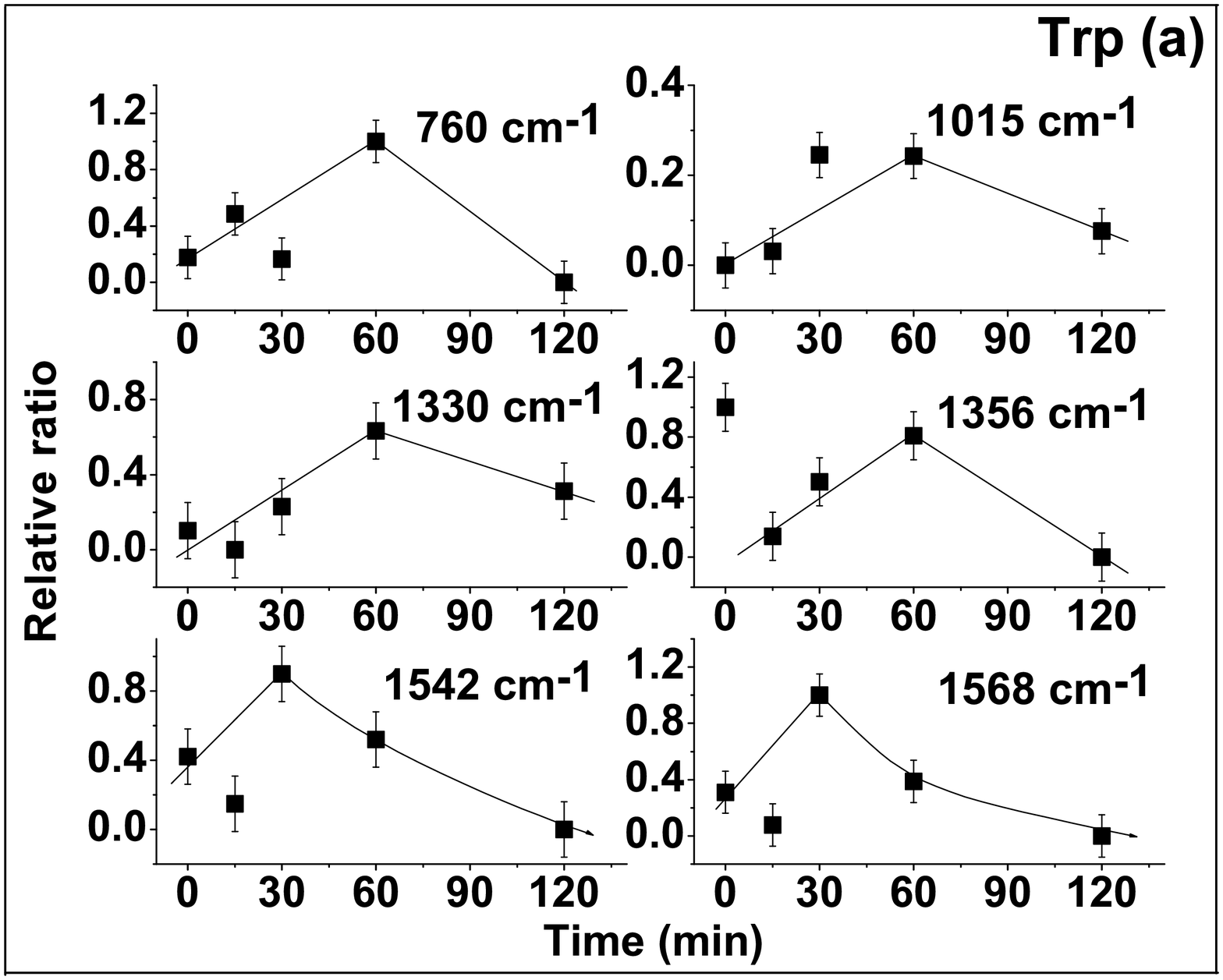}}
\epsfxsize=2.5in\epsffile{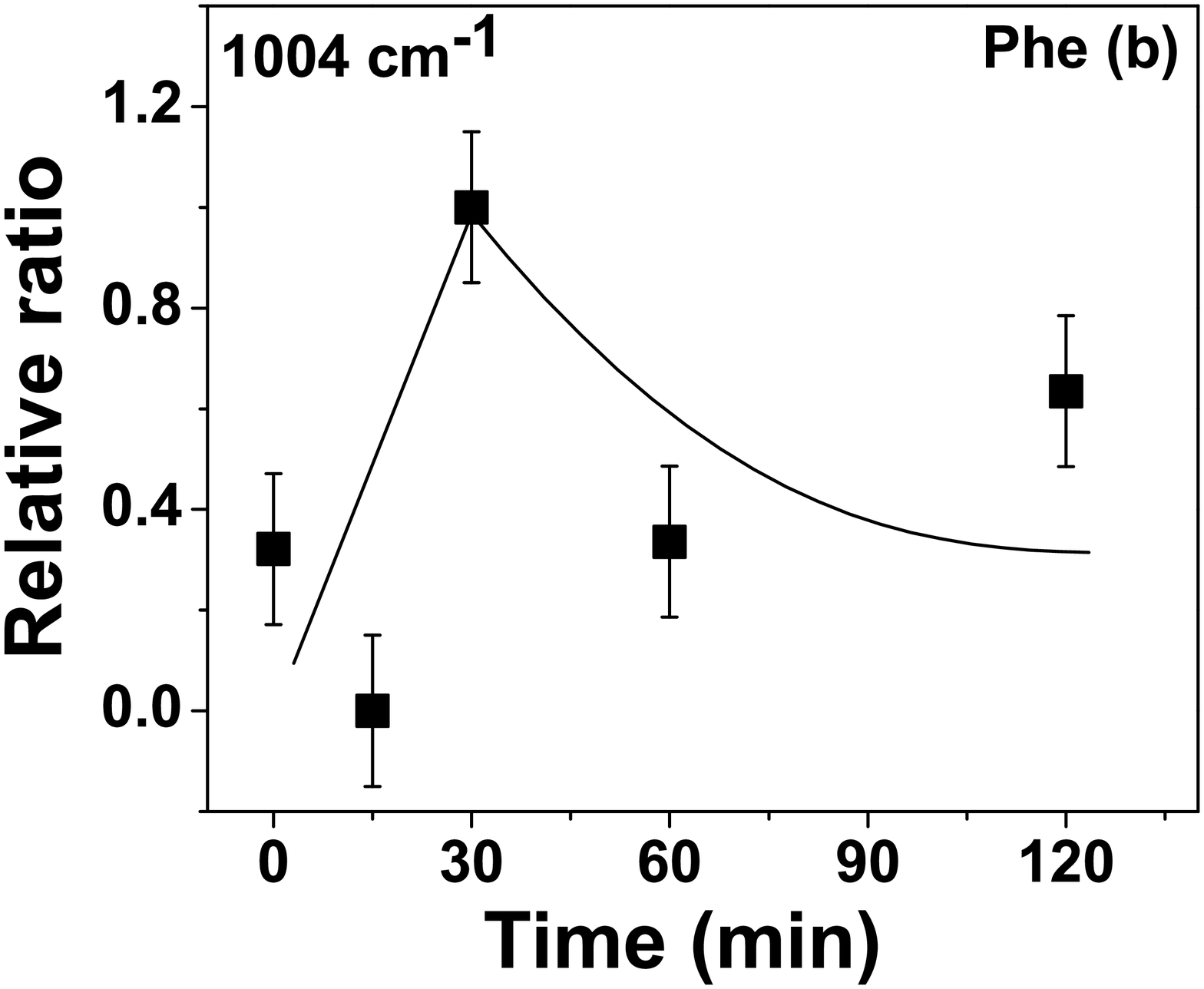}
\epsfxsize=2.5in\epsffile{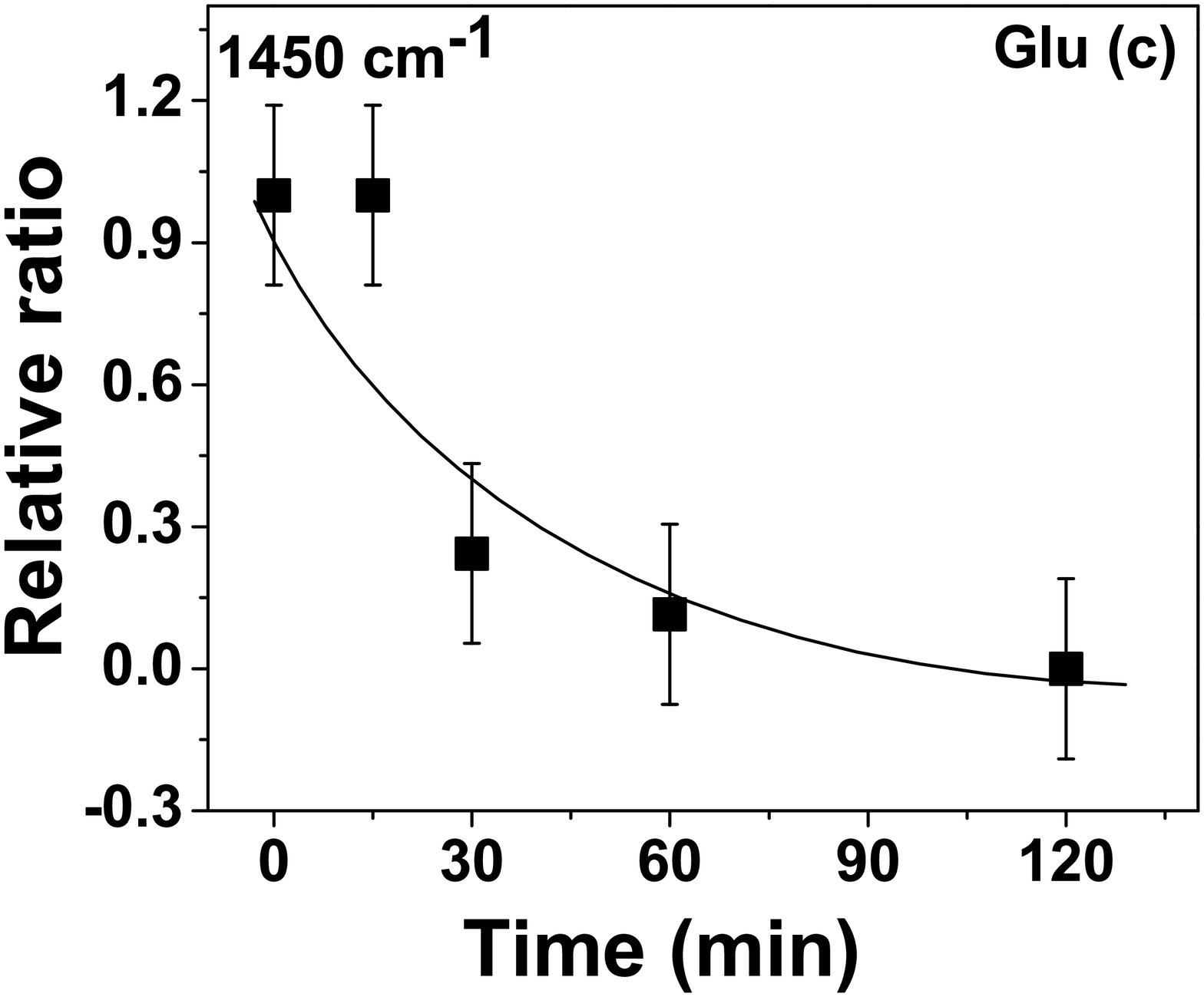}
\epsfxsize=2.5in\epsffile{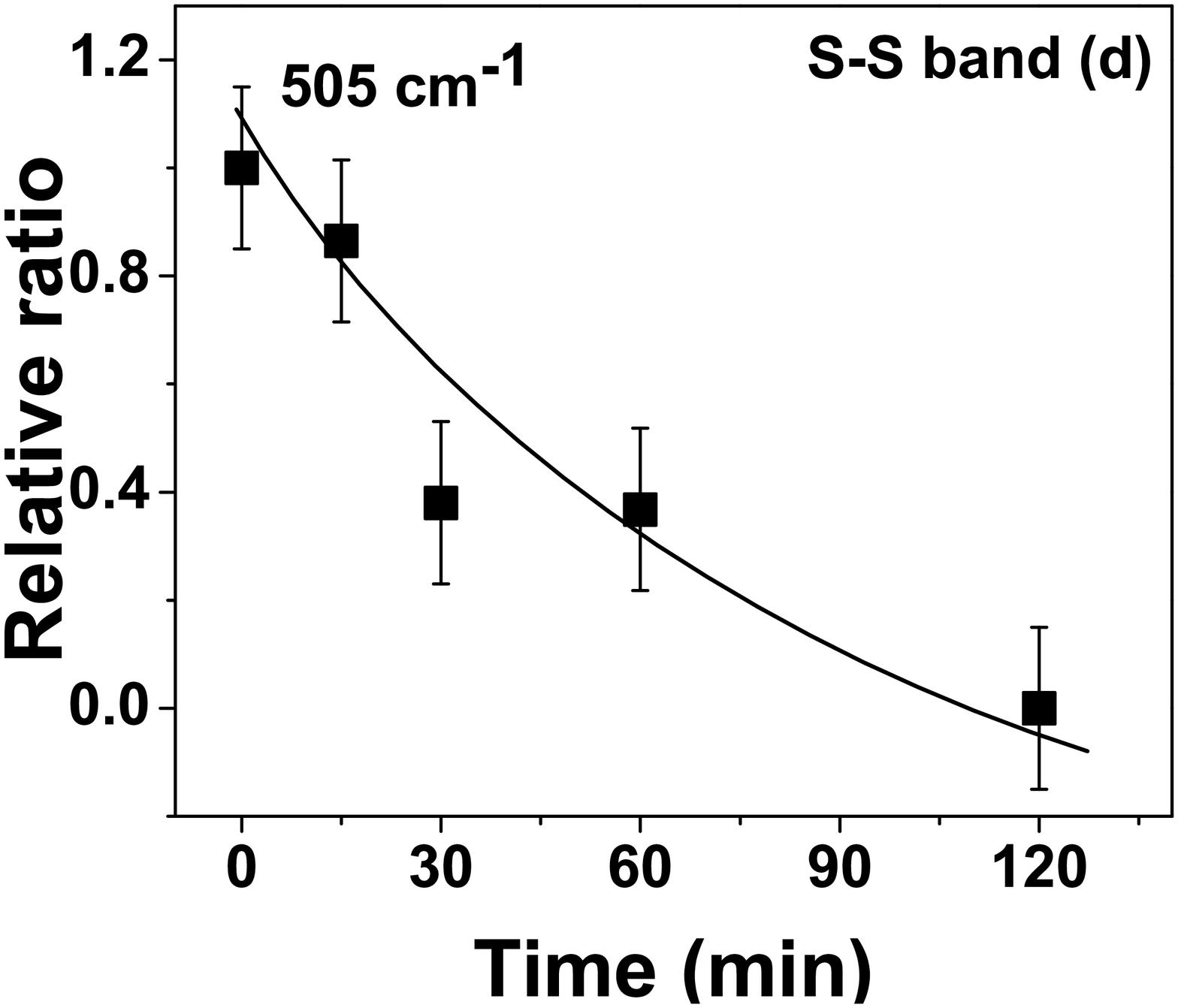}
\epsfxsize=2.5in\epsffile{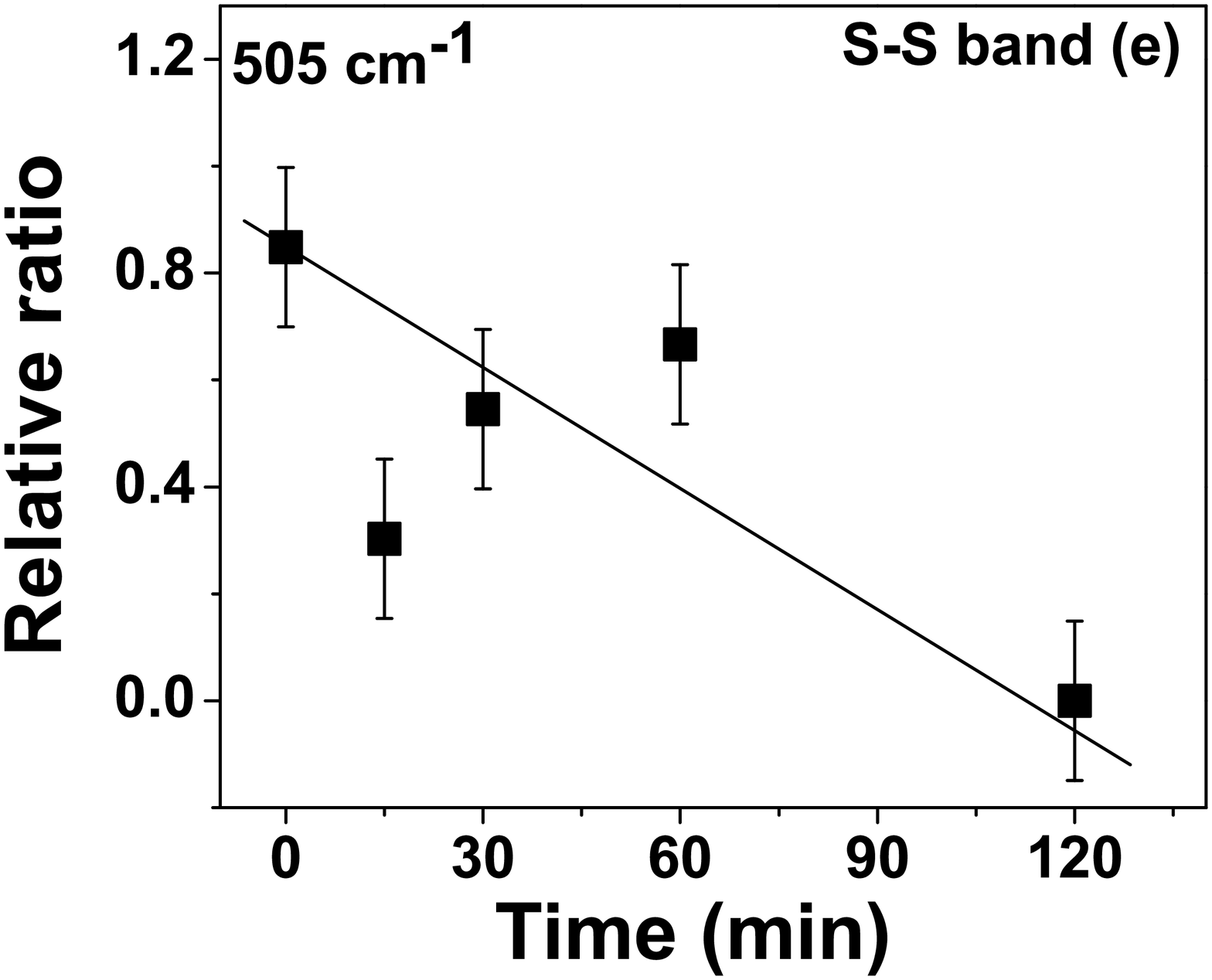}
\vspace*{0.75in}\caption{Chandra et al} \label{Trp}
\end{figure}

\newpage




\newpage

\begin{figure}
\centerline{\epsfxsize=2.5in\epsffile{lys12.eps}}
\vspace*{0.75in}\caption{Chandra et al} \label{cartoon1}
\end{figure}

\vspace*{4.0in}

\newpage
\begin{table}
\caption{Chandra et al}
\begin{tabular}{|c|c|c|} \hline
Assignment  &  peak (cm $^{-1}$)& Reference \\ \hline
S-S   & 505 & \cite{Hu,Lord}\\
Phe & 620 & \cite{Chen} \\
Tyr & 644 & \cite{Chen} \\
Trp - indole breathing (W18)  &  763 & \cite{Chuang}\\
Trp- N$_1$H site vibration (W17) &  877 &\cite{Chuang}  \\
Phe - ring breathing & 1004 &\cite{Hu} \\
Trp - ring breathing & 1015 & \cite{Hu} \\
Trp- indole ring (C-H deformation)  & 1338 & \cite{Chuang}  \\
Trp- W7 wagging  & 1356 &\cite{Chuang}\\
$\delta$(CH$_2$) & 1442& \cite{Chen}\\
Glu-$\delta$(CH$_2$) & 1449&\cite{Navarrete}\\
Trp -indole ring stretching   & 1540 &\cite{Chuang} \\
Trp -W2 mode  & 1568 &\cite{Chuang}\\
Phe/Tyr & 1598 & \cite{Hu}\\
Amide - I  & 1644 & \cite{Hu}\\
    \hline
\end{tabular}
\label{SERST}
\end{table}
\vspace*{4.5in}

\begin{table}
\caption{Chandra et al}
\begin{tabular}{|c|c|c|c|} \hline
Trp sequence  &  ASA (\AA $^{2}$)&Phe-sequence & ASA (\AA $^{2}$)\\
\hline
Trp-28 & 0.12 & Phe-3 &17.82\\
Trp-62 & 120.57 & Phe-34 &64.02\\
Trp-63 & 48.01 &Phe-38 & 15.33\\
Trp-108 & 10.59 &Glu-7 &82.44\\
Trp-111 & 12.69 &Glu-35 &34.16\\
Trp-123 & 55.5 & &\\
    \hline
\end{tabular}
\label{SERS}
\end{table}

\end{document}